 \documentclass [12pt,a4paper,     ]{article}

\usepackage{times}

\DeclareFontFamily{OT1}{times}{}
\DeclareFontShape {OT1}{times}{m }{n }{ <-> ptmr }{}
\DeclareFontShape {OT1}{times}{bx}{n }{ <-> ptmb }{}
\DeclareFontShape {OT1}{times}{m }{it}{ <-> ptmri}{}
\DeclareFontShape {OT1}{times}{bx}{it}{ <-> ptmbi}{}
\usepackage{amsmath}
\usepackage{amsfonts}
\usepackage{amssymb}
\usepackage{latexsym}
\newcommand{\cl}{C \kern -0.1em \ell} 

\newcommand{\CON}{\overline}          

\newcommand{\Scal}{\mathbb{S}}        

\newcommand{\VEC}{\vec{\kern +.1em[}} 
\newcommand{\TOR}{\vec{\kern +.2em]}} 
\newcommand{\BRA}{\langle\kern -.2em\langle} 
\newcommand{\KET}{\rangle\kern -.2em\rangle} 

\begin{document}

\title{\bf\vspace{-1.0cm} Integer-quaternion formulation of Lambek's representation of fundamental particles and their interactions}

\author{Andre Gsponer\\ ~\\ 
\emph{Independent Scientific Research Institute}\\ 
\emph{Box 30, CH-1211 Geneva-12, Switzerland}}

\date{ISRI-02-03.9 ~~ \today}

\maketitle

\begin{abstract}
Lambek's unified classification of the elementary interaction-quanta of the ``Standard model'' is formulated in terms of the 24 units of the integer-quaternion ring, i.e., the tetrahedral group $Q_{24}$. An extension of Lambek's scheme to the octahedral group $Q_{48}$ may enable to take all three generations of leptons and quarks into account, as well as to provide a quantitative explanation for flavor-mixing.
\end{abstract}

\section{Lambek's quaternion assignment to fundamental particles and interactions}
\label{sec:1}

In 2000, Joachim Lambek proposed a remarkable unified classification of elementary interaction quanta, i.e., particles, in which the four first-generation leptons and quarks (i.e., the electron and neutrino, and the up and down quarks) are treated on the same footing, together with their antiparticles and the gauge bosons responsible for their weak, electromagnetic, and strong interactions \cite{LAMBE2000-}.

While the problem of classifying elementary particles is not new, and numerous studies have been published on the possibility of using the quaternion algebra to explain their properties, e.g., Refs.~\cite{LAMBE1950-, GURSE1958A, DEBRO1963A, DEBRO1963B, LAMBE1996-, BEIL-2003-}, Lambek's idea provides a unique classification scheme for the 24 fundamental quantum-number-changing quanta of the current ``Standard model.'' 

Indeed, the novelty of Lambek approach is to assemble in a single multiplet the 16 elementary fermions together with the 8 gauge bosons which modify the quantum numbers of these fermions during interactions.\footnote{The 4 ``neutral'' gauge bosons (i.e., the photon, the $Z^0$, and the two gluons which do not change color) are identified with zero.}  The resulting multiplet is then associated with 24 quaternions whose components (equal to $+1$, $0$, or $-1$) uniquely identify each member (see \emph{Table~\ref{tbl:1}}).  Moreover, when considering the set of all possible Feynman vertices in which these particles are involved, Lambek found that adding or subtracting these quaternions insures that all quantum numbers are properly conserved during interaction.

In \emph{Table~\ref{tbl:1}}, the scalar part $F_{\text{nb}}$ of the quaternion $\Lambda$ associated with each particle, i.e.,
\begin{equation}\label{equ:1}
           F_{\text{nb}}  = \Scal [ \Lambda ] ,
\end{equation}
which Lambek calls the ``Fermion number,'' generalizes the concepts of leptonic and baryonic numbers.  The electric charge $Z_{\text{el}}$ of the particle is determined by the ``trace'' of the vector part of  $\Lambda$, i.e., the sum of its components, which is obtained by addition, or by calculating the scalar product
\begin{equation}\label{equ:2}
Z_{\text{el}} = \tfrac{1}{3} \Scal [ \vec{e} \CON{\Lambda} ] ,
\end{equation}
that projects the quaternion $\Lambda$ on the vector
\begin{equation}\label{equ:3}
\vec{e}= i + j + k ,
\end{equation}
which is just the sum of the three non-scalar units defining Hamilton's quaternions.

\emph{Table~\ref{tbl:1}} also gives the values of the standard ``baryon number'' $N$  and ``isospin projection'' $I_z$.  The electric charge is obtained  from these numbers by the celebrated 1932 ``Heisenberg formula:''
 \begin{equation}\label{equ:4}
Z_{\text{el}} = \tfrac{1}{2}N \pm I_z ,
\end{equation}
which was generalized in 1953-54 by Gell-Mann and Nishijima to account for strangeness (and later by others for charm and beauty) in order to accommodate all baryons and mesons. In a form or another, formula \eqref{equ:4} is a fundamental requirement for all classifications and models of elementary particles.

An interesting point addressed in Lambek's paper is the question why there are only 24 actual particles when there are in total 81 quaternions with components equal to $+1$, $0$, or $-1$.  Looking at  \emph{Table~\ref{tbl:1}}, Lambek observed the following empirical rule: \emph{The number of positive entries in the associated quaternion is either zero or odd, and ditto for the number of negative entries}. This rule has an interesting consequence: \emph{The quaternion conjugate $\CON{\Lambda}$ of the fermionic entries do not correspond to actual particles. (Except for the neutrinos which are self-conjugate.)}  Obviously, this observation is just the consequence of the fact that applying quaternion conjugation to all fermions simply reverses the sign of all electric charges, which is equivalent to redefining the conventional overall sign of electric charge. The possibility of this redefinition reduces the total number of fundamental quaternions from 81 to about 40.  However, there is no obvious reason why, for example, $1+i-j$ should not correspond to a particle when $1-i-j$ does. It is therefore of interest to see whether there could be a mathematical explanation for the limitation of the number of fundamental particles to just 24, excluding the four ``neutral'' bosons which correspond to zero in this classification.\footnote{~ $\gamma$ and $Z^0$, as well as $g_{\CON{C}C}$ and $g_{C\CON{C}}$, are not \emph{strictly} neutral since they couple to either electro-weakly, or strongly, interacting particles.}

\section{Hurwitz's integer quaternions}
\label{sec:2}

The facts that Lambek's representation involves quaternions whose components are integers, and that just 24 of these are required to fully describe the first generation of elementary particles and their interactions, suggest that there could be a connection between Lambek's classification and Adolf Hurwitz's mathematical theory of integer quaternions \cite{HURWI1919-,HURWI1963-}.

According to Hurwitz, a quaternion
\begin{equation}\label{equ:5}
q = w + x i + y j + z k ,
\end{equation}
is called \emph{integer}, if the coefficients $w, x, y, z$ are either all integers or all of the form $n + \frac{1}{2}$.  In other words, integer quaternions are linear combinations of
\begin{equation}\label{equ:6}
i, ~~ j, ~~ k, ~~ \frac{(1+i+j+k)}{2} .
\end{equation} 
Integer quaternions form a ring. They have the fundamental property that while there is an infinite number of real quaternions $r$ with unit norm, i.e., such that $r\CON{r}=1$, there are only 24 integer quaternions $q$ such that $q\CON{q}=1$. These \emph{integer quaternion units}, or \emph{Hurwitz units} $\{H_n\}$ to distinguish them from the \emph{Hamilton units} $\{\pm{e_n}\}$, are 
\begin{equation}\label{equ:7}
\pm{1}, ~ \pm{i}, ~ \pm{j}, ~ \pm{k}, ~\frac{(\pm{1}+\pm{i}+\pm{j}+\pm{k})}{2} .
\end{equation}
They are displayed in \emph{Table~\ref{tbl:2}} together with the quantum numbers derived with the same rules as in \emph{Table~\ref{tbl:1}}.

The integer quaternion units are related to the finite quaternion groups that were classified in 1881 by W.I.\ Stringham \cite{STRIN1881-}.  These groups, which comprise the quaternions whose products and powers are also quaternions of the same group, consist of two series, the cyclic and dihedral groups $C_n$ and $D_n$, and of the tetrahedral,\footnote{As an example of a different application of tetrahedral symmetry in elementary particle physics, see J.S.R.\ Chisholm and R.S.\ Farwell, \emph{Tetrahedral structure of idempotents of the Clifford algebra $\cl_{1,3}$}, in: A. Micali et al., eds., Clifford Algebras and their Applications in Mathematical Physics (Kluwer Academic Publishers, Dordrecht, 1992) 27--32.} octahedral, and icosahedral groups $Q_{24}$, $Q_{48}$, $Q_{120}$ --- which are of order 2$n$, 4$n$, 24, 48, and 120, respectively. These five groups, combined to spatial rotations and reflections, generate the 32 symmetry groups of crystallography. 

Hurwitz's units are identical to Stringham's group $Q_{24}$, and Hamilton's units $\{\pm{e_n}\} = \{\pm 1,\pm i,\pm j,\pm k\}$ form a permutable subgroup $Q_{8}$ of $Q_{24}$.\footnote{$Q_{8}$ and $Q_{24}$ are contained in $Q_{48}$ as permutable subgroups. However, the group $Q_{120}$ contains no permutable subgroup except the group $\{\pm 1\}$ of order 2.}

\section{Lambek's quaternion assignment expressed with Hurwitz units}
\label{sec:3}

There are many possible ways to express Lambek's quaternions in terms of Hurwitz units.  However, an essentially unique one is provided by postulating that Lambek's quaternions correspond to a 4-dimensional generalization of charge, and that these \emph{Lambek charges} $\Lambda$ are expressible as isospin-doublets in terms of Hurwitz units $H_n \in Q_{24}$ by a simple generalization of Heisenberg's formula \eqref{equ:4}, i.e.,
\begin{gather}
    \Lambda_{(+)} = H_n + H_m       \label{equ:8},\\
    \Lambda_{(-)} = H_n + \CON{H_m} \label{equ:9},
\end{gather}
where quaternion conjugation is used to reverse the sign of the ``isospin projection.''

Strictly speaking, the Heisenberg--Gell-Mann--Nishijima formula applies only to hadrons.  In practice, however, it also applies to leptons if we merge the baryon and lepton quantum numbers in a single fermion number applicable to both leptons and quarks.  As shown in \emph{Table~\ref{tbl:3}}, we get then the following assignments for the elementary fermions:
\begin{gather}\label{equ:10}
    \nu_0 =  +h_8+     h_1  ,  \\
    e^-   =  +h_8+\CON{h_1} ,  \\
    u_R   =  +h_5+     h_1  ,  \\
    d_R   =  +h_5+\CON{h_1} ,  \\
    u_B   =  +h_3+     h_1  ,  \\
    d_B   =  +h_3+\CON{h_1} ,  \\
    u_G   =  +h_2+     h_1  ,  \\
    d_G   =  +h_2+\CON{h_1} . 
\end{gather}
Because $h_8 = \CON{h_1}$, these expressions can be rewritten using $\CON{h_8}$ instead of $h_1$ on the right hand side.  Moreover, since $\nu_0 =  +h_8+\CON{h_8} = 1$, we can put 1 for $\nu_0$ in \emph{Table~\ref{tbl:3}}. The corresponding antifermions are obtained by just changing the plus signs into minus signs.

It is surprising at first that these expressions do not involve $h_4$, $h_6$, and $h_7$.  In fact, if we use these units according to Eqs.~\eqref{equ:8} and \eqref{equ:9} we find:
\begin{gather}\label{equ:11}
    h_4+      h_1  = \CON{d_R} , \\
    h_4+ \CON{h_1} = \CON{u_R} , \\
    h_6+      h_1  = \CON{d_B} , \\
    h_6+ \CON{h_1} = \CON{u_B} , \\
    h_7+      h_1  = \CON{d_G} , \\
    h_7+ \CON{h_1} = \CON{u_G} ,
\end{gather}
which are just the quaternion conjugate of the above combinations. Therefore, we can exclude them because they correspond to a redefinition of the overall sign of electric charge.

For the gauge bosons there is no rule such as the Heisenberg--Gell-Mann--Nishijima formula.  However, it seems natural that their Lambek charges can be expressed in terms of Hurwitz's units in a way similar to fermions. Indeed, referring to \emph{Table~\ref{tbl:1}},
\begin{equation}\label{equ:12}
       W^\pm = \pm (h_1 - h_8) = \pm (e_1 + e_2 + e_3) = \pm \vec{e} ,
\end{equation}
i.e., $W^- = \CON{W^+}$ so that $W^\pm$ has isospin $\pm 1$, as it should be.

For the gluons, there appears to be a number of possibilities:
\begin{gather}\label{equ:13}
   g_{\CON{B}G} = +e_2-e_3 =  h_2 - h_3 =  h_6 - h_7  , \\
   g_{\CON{G}R} = -e_1+e_3 =  h_5 - h_2 =  h_7 - h_4  , \\
   g_{\CON{R}B} = +e_1-e_2 =  h_3 - h_5 =  h_4 - h_6  . 
\end{gather}
Because the gluons are vectors, their quaternion conjugates are just their antiparticles.  Since we have not used $h_4$, $h_6$, and $h_7$ for either the fermions or the two charged intermediate vector bosons, we use these units to express the gluons's Lambek charges in terms of Hurwitz units.

\section{Discussion}
\label{sec:4}

As can be seen in \emph{Table~\ref{tbl:3}}, it is possible to formulate Lambek's representation of elementary particles in a way that uses at least once all 24 Hurwitz units.  However, it is not a simple ``one to one'' correspondence, although it can be justified by reference to the Heisenberg--Gell-Mann--Nishijima formula.  Nevertheless, the scheme is attractive, and possibly more promising than the one based on ``integer complex numbers'' which was proposed in reference \cite{GSPON1994A}.\footnote{``Integer complex numbers'' are elements of the rings $R_4 = \{n+m\exp{(i\pi/2)}\}$, or\\ $R_6 = \{n+m\exp{(i\pi/3)}\}$, where $n,m \in \mathbb{Z}$. In this notation, integer quaternions form\\ the ring $R_{24} = \{n_k e_k + m_l h_l\} = Q_{24}$, where $n_l,m_k \in \mathbb{Z}$.}

Concerning the 32 second and third generation quarks and leptons, the question is how to extend Lambek's classification from 24 to 24+32=56 elementary quanta.  Since these 32 extra fermions are related in their interactions by flavor-mixing constraints, a solution could be to extend Lambek's scheme from $Q_{24}$ to the octahedral group $Q_{48}$, a possibility that deserves further investigation.\footnote{With three generations of massive leptons and quarks, the mass-mixing matrices correspond to $2 \times 4 = 8 = 56 - 48$ constraints.}   If this turns out to be correct, one would have a remarkable analogy between the lattice structure of quantized fields and crystallography.

\section*{Acknowledgment}

The author thanks Professor Lambek for inspiring correspondence and helpful comments.
\label{ack}

\newpage

\begin{table}
\begin{center}
\vspace{-1.0cm}
\renewcommand{\arraystretch}{1.2}
\begin{tabular}{|c| rl          rl          rl          rl    |   c    |   c    | c | c |}
\hline
\multicolumn{13}{|c|}{\raisebox{+0.4em}
{{\bf Lambek's quaternion assignments to fundamental particles \rule{0mm}{6mm}}}} \\
\hline
\hline
               &     &     &     &     &     &     &     &     &        &                &   &     \\
Particle       &  &$\Lambda~($&$1,$&$ i,$&$ j,$&$ k$&$)~$&     &$F_{\text{nb}}$&$Z_{\text{el}}$        &$N$&$I_z$\\
               &     &     &     &     &     &     &     &     &        &                &   &     \\
\hline
Gauge bosons   &     &     &     &     &     &     &     &     &        &                &   &     \\
\hline
$\gamma, Z^0  $&$   $&$ 0 $&$   $&$   $&$   $&$   $&$   $&$   $&$   0  $&$   0          $& 0 & 0   \\
$g_{\CON{C}C},g_{C\CON{C}}$&$   $&$ 0 $&$   $&$   $&$   $&$   $&$   $&$   $&$ 0 $&$ 0 $& 0 & 0   \\
\hline

$W^-          $&$   $&$   $&$ - $&$ i $&$ - $&$ j $&$ - $&$ k $&$   0  $&$   -1        $& 0 &$-1$  \\
$W^+          $&$   $&$   $&$ + $&$ i $&$ + $&$ j $&$ + $&$ k $&$   0  $&$   +1        $& 0 &$+1$  \\

$g_{\CON{B}G}$&$   $&$   $&$   $&$   $&$ + $&$ j $&$ - $&$ k $&$   0  $&$   0         $& 0 & 0    \\
$g_{\CON{G}R}$&$   $&$   $&$ - $&$ i $&$   $&$   $&$ + $&$ k $&$   0  $&$   0         $& 0 & 0    \\
$g_{\CON{R}B}$&$   $&$   $&$ + $&$ i $&$ - $&$ j $&$   $&$   $&$   0  $&$   0         $& 0 & 0    \\

$g_{\CON{G}B}$&$   $&$   $&$   $&$   $&$ - $&$ j $&$ + $&$ k $&$   0  $&$   0         $& 0 & 0    \\
$g_{\CON{R}G}$&$   $&$   $&$ + $&$ i $&$   $&$   $&$ - $&$ k $&$   0  $&$   0         $& 0 & 0    \\
$g_{\CON{B}R}$&$   $&$   $&$ - $&$ i $&$ + $&$ j $&$   $&$   $&$   0  $&$   0         $& 0 & 0    \\
\hline

Fermions       &     &     &     &     &     &     &     &     &        &               &   &      \\
\hline
$\nu_0         $&$ + $&$ 1 $&$   $&$   $&$   $&$   $&$   $&$   $&$  +1  $&$         0  $&$-1          $&$+\frac{1}{2}$\\
$e^-           $&$ + $&$ 1 $&$ - $&$ i $&$ - $&$ j $&$ - $&$ k $&$  +1  $&$        -1  $&$-1          $&$-\frac{1}{2}$\\

$u_R           $&$ + $&$ 1 $&$   $&$   $&$ + $&$ j $&$ + $&$ k $&$  +1  $&$+\frac{2}{3}$&$+\frac{1}{3}$&$+\frac{1}{2}$\\
$u_B           $&$ + $&$ 1 $&$ + $&$ i $&$   $&$   $&$ + $&$ k $&$  +1  $&$+\frac{2}{3}$&$+\frac{1}{3}$&$+\frac{1}{2}$\\
$u_G           $&$ + $&$ 1 $&$ + $&$ i $&$ + $&$ j $&$   $&$   $&$  +1  $&$+\frac{2}{3}$&$+\frac{1}{3}$&$+\frac{1}{2}$\\

$d_R           $&$ + $&$ 1 $&$ - $&$ i $&$   $&$   $&$   $&$ k $&$  +1  $&$-\frac{1}{3}$&$+\frac{1}{3}$&$-\frac{1}{2}$\\
$d_B           $&$ + $&$ 1 $&$   $&$   $&$ - $&$ j $&$   $&$   $&$  +1  $&$-\frac{1}{3}$&$+\frac{1}{3}$&$-\frac{1}{2}$\\
$d_G           $&$ + $&$ 1 $&$   $&$   $&$   $&$   $&$ - $&$ k $&$  +1  $&$-\frac{1}{3}$&$+\frac{1}{3}$&$-\frac{1}{2}$\\
\hline

Antifermions    &     &     &     &     &     &     &     &     &        &              &   &      \\
\hline
$\nu_{\CON{0}}$&$ - $&$ 1 $&$   $&$   $&$   $&$   $&$   $&$   $&$  -1  $&$         0  $&$+1          $&$-\frac{1}{2}$\\
$e^+           $&$ - $&$ 1 $&$ + $&$ i $&$ + $&$ j $&$ + $&$ k $&$  -1  $&$        +1  $&$+1          $&$+\frac{1}{2}$\\

$u_{\CON{R}}  $&$ - $&$ 1 $&$   $&$   $&$ - $&$ j $&$ - $&$ k $&$  -1  $&$-\frac{2}{3}$&$-\frac{1}{3}$&$-\frac{1}{2}$\\
$u_{\CON{B}}  $&$ - $&$ 1 $&$ - $&$ i $&$   $&$   $&$ - $&$ k $&$  -1  $&$-\frac{2}{3}$&$-\frac{1}{3}$&$-\frac{1}{2}$\\
$u_{\CON{G}}  $&$ - $&$ 1 $&$ - $&$ i $&$ - $&$ j $&$   $&$   $&$  -1  $&$-\frac{2}{3}$&$-\frac{1}{3}$&$-\frac{1}{2}$\\

$d_{\CON{R}}  $&$ - $&$ 1 $&$ + $&$ i $&$   $&$   $&$   $&$ k $&$  -1  $&$+\frac{1}{3}$&$-\frac{1}{3}$&$+\frac{1}{2}$\\
$d_{\CON{B}}  $&$ - $&$ 1 $&$   $&$   $&$ + $&$ j $&$   $&$   $&$  -1  $&$+\frac{1}{3}$&$-\frac{1}{3}$&$+\frac{1}{2}$\\
$d_{\CON{G}}  $&$ - $&$ 1 $&$   $&$   $&$   $&$   $&$ + $&$ k $&$  -1  $&$+\frac{1}{3}$&$-\frac{1}{3}$&$+\frac{1}{2}$\\

\hline
\end{tabular}
\end{center}
\caption{Lambek's quaternion assignments to fundamental particles.}
\label{tbl:1}
\renewcommand{\arraystretch}{1.0}
\end{table}

\newpage

\begin{table}
\begin{center}
\renewcommand{\arraystretch}{1.2}
\begin{tabular}{|c|rl          rl          rl          rl    |   c    |   c    | }
\hline
\multicolumn{11}{|c|}{\raisebox{+0.4em}
{{\bf Hurwitz's integer-quaternion units \rule{0mm}{6mm}}}} \\
 \hline
\hline
             &     &     &     &     &     &     &     &     &        &                 \\
Units &     &     &     &     &     &     &     &     &$F_{\text{nb}}$&$Z_{\text{el}}$  \\
             &     &     &     &     &     &     &     &     &        &                 \\
\hline
$+e_0       $&$ + $&$ 1 $&$   $&$   $&$   $&$   $&$   $&$   $&$  +1  $&$   0   $        \\
$-e_0       $&$ - $&$ 1 $&$   $&$   $&$   $&$   $&$   $&$   $&$  -1  $&$   0   $        \\
$+e_1       $&$   $&$   $&$ + $&$ i $&$   $&$   $&$   $&$   $&$   0  $&$+\frac{1}{3}$   \\
$-e_1       $&$   $&$   $&$ - $&$ i $&$   $&$   $&$   $&$   $&$   0  $&$-\frac{1}{3}$   \\
$+e_2       $&$   $&$   $&$   $&$   $&$ + $&$ j $&$   $&$   $&$   0  $&$+\frac{1}{3}$   \\
$-e_2       $&$   $&$   $&$   $&$   $&$ - $&$ j $&$   $&$   $&$   0  $&$-\frac{1}{3}$   \\
$+e_3       $&$   $&$   $&$   $&$   $&$   $&$   $&$ + $&$ k $&$   0  $&$+\frac{1}{3}$   \\
$-e_3       $&$   $&$   $&$   $&$   $&$   $&$   $&$ - $&$ k $&$   0  $&$-\frac{1}{3}$   \\
\hline
$+h_1       $&$ + $&$\frac{1}{2}$&$ + $&$\frac{1}{2}i $&$ + $&$\frac{1}{2}j $&$ + $&$\frac{1}{2}k $&$+\frac{1}{2}$&$+\frac{1}{2}$   \\
$+h_2       $&$ + $&$\frac{1}{2}$&$ + $&$\frac{1}{2}i $&$ + $&$\frac{1}{2}j $&$ - $&$\frac{1}{2}k $&$+\frac{1}{2}$&$+\frac{1}{6}$   \\
$+h_3       $&$ + $&$\frac{1}{2}$&$ + $&$\frac{1}{2}i $&$ - $&$\frac{1}{2}j $&$ + $&$\frac{1}{2}k $&$+\frac{1}{2}$&$+\frac{1}{6}$   \\
$+h_4       $&$ + $&$\frac{1}{2}$&$ + $&$\frac{1}{2}i $&$ - $&$\frac{1}{2}j $&$ - $&$\frac{1}{2}k $&$+\frac{1}{2}$&$-\frac{1}{6}$   \\
$+h_5       $&$ + $&$\frac{1}{2}$&$ - $&$\frac{1}{2}i $&$ + $&$\frac{1}{2}j $&$ + $&$\frac{1}{2}k $&$+\frac{1}{2}$&$+\frac{1}{6}$   \\
$+h_6       $&$ + $&$\frac{1}{2}$&$ - $&$\frac{1}{2}i $&$ + $&$\frac{1}{2}j $&$ - $&$\frac{1}{2}k $&$+\frac{1}{2}$&$-\frac{1}{6}$   \\
$+h_7       $&$ + $&$\frac{1}{2}$&$ - $&$\frac{1}{2}i $&$ - $&$\frac{1}{2}j $&$ + $&$\frac{1}{2}k $&$+\frac{1}{2}$&$-\frac{1}{6}$   \\
$+h_8       $&$ + $&$\frac{1}{2}$&$ - $&$\frac{1}{2}i $&$ - $&$\frac{1}{2}j $&$ - $&$\frac{1}{2}k $&$+\frac{1}{2}$&$-\frac{1}{2}$   \\
\hline
$-h_8       $&$ - $&$\frac{1}{2}$&$ + $&$\frac{1}{2}i $&$ + $&$\frac{1}{2}j $&$ + $&$\frac{1}{2}k $&$-\frac{1}{2}$&$+\frac{1}{2}$   \\
$-h_7       $&$ - $&$\frac{1}{2}$&$ + $&$\frac{1}{2}i $&$ + $&$\frac{1}{2}j $&$ - $&$\frac{1}{2}k $&$-\frac{1}{2}$&$+\frac{1}{6}$   \\
$-h_6       $&$ - $&$\frac{1}{2}$&$ + $&$\frac{1}{2}i $&$ - $&$\frac{1}{2}j $&$ + $&$\frac{1}{2}k $&$-\frac{1}{2}$&$+\frac{1}{6}$   \\
$-h_5       $&$ - $&$\frac{1}{2}$&$ + $&$\frac{1}{2}i $&$ - $&$\frac{1}{2}j $&$ - $&$\frac{1}{2}k $&$-\frac{1}{2}$&$-\frac{1}{6}$   \\
$-h_4       $&$ - $&$\frac{1}{2}$&$ - $&$\frac{1}{2}i $&$ + $&$\frac{1}{2}j $&$ + $&$\frac{1}{2}k $&$-\frac{1}{2}$&$+\frac{1}{6}$   \\
$-h_3       $&$ - $&$\frac{1}{2}$&$ - $&$\frac{1}{2}i $&$ + $&$\frac{1}{2}j $&$ - $&$\frac{1}{2}k $&$-\frac{1}{2}$&$-\frac{1}{6}$   \\
$-h_2       $&$ - $&$\frac{1}{2}$&$ - $&$\frac{1}{2}i $&$ - $&$\frac{1}{2}j $&$ + $&$\frac{1}{2}k $&$-\frac{1}{2}$&$-\frac{1}{6}$   \\
$-h_1       $&$ - $&$\frac{1}{2}$&$ - $&$\frac{1}{2}i $&$ - $&$\tfrac{1}{2}j$&$ - $&$\frac{1}{2}k $&$-\frac{1}{2}$&$-\frac{1}{2}$   \\
\hline
\end{tabular}
\end{center}
\caption{Hurwitz units.}
\label{tbl:2}
\renewcommand{\arraystretch}{1.0}
\end{table}

\newpage

\begin{table}
\begin{center}
\vspace{-1.0cm}
\renewcommand{\arraystretch}{1.2}
\begin{tabular}{|c| rl          rl          rl          rl    |   c    |   c    | c | }
\hline
\multicolumn{12}{|c|}{\raisebox{+0.4em}
{{\bf Lambek's quaternion-charge assignments in terms of Hurwitz units \rule{0mm}{6mm}}}} \\
\hline
\hline
              &     &     &     &     &     &     &     &     &        &               &                   \\
Particle      &  &$\Lambda~($&$1,$&$ i,$&$ j,$&$ k$&$)~$&  &$F_{\text{nb}}$&$Z_{\text{el}}$    &$\Lambda~(~~e_n,~~h_m~~)$\\
              &     &     &     &     &     &     &     &     &        &               &                   \\
\hline
Gauge bosons  &     &     &     &     &     &     &     &     &        &               &                   \\
\hline
$\gamma, Z^0                $&$ ~ $&$ 0 $&$ $&$ $&$ $&$ $&$ $&$ $&$ 0  $&$   0   $     &                   \\
$g_{\CON{C}C},g_{C\CON{C}}$&$ ~ $&$ 0 $&$ $&$ $&$ $&$ $&$ $&$ $&$ 0  $&$   0   $     &                   \\
\hline

$W^-          $&$   $&$   $&$ - $&$ i $&$ - $&$ j $&$ - $&$ k $&$   0  $&$   -1  $     & $+e_1+e_2+e_3$    \\
$W^+          $&$   $&$   $&$ + $&$ i $&$ + $&$ j $&$ + $&$ k $&$   0  $&$   +1  $     & $-e_1-e_2-e_3$    \\

$g_{\CON{B}G}$&$   $&$   $&$   $&$   $&$ + $&$ j $&$ - $&$ k $&$   0  $&$   0   $     & $+h_6-h_7$        \\
$g_{\CON{G}R}$&$   $&$   $&$ - $&$ i $&$   $&$   $&$ + $&$ k $&$   0  $&$   0   $     & $+h_7-h_4$        \\
$g_{\CON{R}B}$&$   $&$   $&$ + $&$ i $&$ - $&$ j $&$   $&$   $&$   0  $&$   0   $     & $+h_4-h_6$        \\

$g_{\CON{G}B}$&$   $&$   $&$   $&$   $&$ - $&$ j $&$ + $&$ k $&$   0  $&$   0   $     & $-h_6+h_7$        \\
$g_{\CON{R}G}$&$   $&$   $&$ + $&$ i $&$   $&$   $&$ - $&$ k $&$   0  $&$   0   $     & $-h_7+h_4$        \\
$g_{\CON{B}R}$&$   $&$   $&$ - $&$ i $&$ + $&$ j $&$   $&$   $&$   0  $&$   0   $     & $-h_4+h_6$        \\
\hline

Fermions      &     &     &     &     &     &     &     &     &        &               &                   \\
\hline
$\nu_0       $&$ + $&$ 1 $&$   $&$   $&$   $&$   $&$   $&$   $&$  +1  $&$         0  $ &$    1          $  \\
$e^-         $&$ + $&$ 1 $&$ - $&$ i $&$ - $&$ j $&$ - $&$ k $&$  +1  $&$        -1  $ &$+h_8+\CON{h_1}$  \\

$u_R         $&$ + $&$ 1 $&$   $&$   $&$ + $&$ j $&$ + $&$ k $&$  +1  $&$+\frac{2}{3}$ &$+h_5+      h_1 $  \\
$u_B         $&$ + $&$ 1 $&$ + $&$ i $&$   $&$   $&$ + $&$ k $&$  +1  $&$+\frac{2}{3}$ &$+h_3+      h_1 $  \\
$u_G         $&$ + $&$ 1 $&$ + $&$ i $&$ + $&$ j $&$   $&$   $&$  +1  $&$+\frac{2}{3}$ &$+h_2+      h_1 $  \\

$d_R         $&$ + $&$ 1 $&$ - $&$ i $&$   $&$   $&$   $&$ k $&$  +1  $&$-\frac{1}{3}$ &$+h_5+\CON{h_1}$  \\
$d_B         $&$ + $&$ 1 $&$   $&$   $&$ - $&$ j $&$   $&$   $&$  +1  $&$-\frac{1}{3}$ &$+h_3+\CON{h_1}$  \\
$d_G         $&$ + $&$ 1 $&$   $&$   $&$   $&$   $&$ - $&$ k $&$  +1  $&$-\frac{1}{3}$ &$+h_2+\CON{h_1}$  \\
\hline

Antifermions   &     &     &     &     &     &     &     &     &        &              &                   \\
\hline
$\nu_{\CON{0}}$&$ - $&$ 1 $&$   $&$   $&$   $&$   $&$   $&$   $&$  -1  $&$        0  $&$    -1         $  \\
$e^+          $&$ - $&$ 1 $&$ + $&$ i $&$ + $&$ j $&$ + $&$ k $&$  -1  $&$        +1  $&$-h_8-\CON{h_1}$  \\

$u_{\CON{R}} $&$ - $&$ 1 $&$   $&$   $&$ - $&$ j $&$ - $&$ k $&$  -1  $&$-\frac{2}{3}$&$-h_5-      h_1 $  \\
$u_{\CON{B}} $&$ - $&$ 1 $&$ - $&$ i $&$   $&$   $&$ - $&$ k $&$  -1  $&$-\frac{2}{3}$&$-h_3-      h_1 $  \\
$u_{\CON{G}} $&$ - $&$ 1 $&$ - $&$ i $&$ - $&$ j $&$   $&$   $&$  -1  $&$-\frac{2}{3}$&$-h_2-      h_1 $  \\

$d_{\CON{R}} $&$ - $&$ 1 $&$ + $&$ i $&$   $&$   $&$   $&$ k $&$  -1  $&$+\frac{1}{3}$&$-h_5-\CON{h_1}$  \\
$d_{\CON{B}} $&$ - $&$ 1 $&$   $&$   $&$ + $&$ j $&$   $&$   $&$  -1  $&$+\frac{1}{3}$&$-h_3-\CON{h_1}$  \\
$d_{\CON{G}} $&$ - $&$ 1 $&$   $&$   $&$   $&$   $&$ + $&$ k $&$  -1  $&$+\frac{1}{3}$&$-h_2-\CON{h_1}$  \\

\hline
\end{tabular}
\end{center}
\caption{Lambek's quaternion-charge assignments in terms of Hurwitz units.}
\label{tbl:3}
\renewcommand{\arraystretch}{1.0}
\end{table}

\end{document}